\documentclass{ws-ijns}
\usepackage{graphicx}
\usepackage[english]{babel}
\usepackage[super,sort,compress]{cite}
\begin{document}


\markboth{Leming, M., Gorriz, J.M., \& Suckling, J.}{Ensemble deep learning on large, mixed-site fMRI datasets in autism and other tasks}

\title{ENSEMBLE DEEP LEARNING ON LARGE, MIXED-SITE FMRI DATASETS IN AUTISM AND OTHER TASKS}

\author{MATTHEW LEMING}

\address{Department of Psychiatry, University of Cambridge, Herchel Smith Building, Robinson Way,\\
Cambridge, CB20SZ, UK\\
E-mail: ml784@cam.ac.uk\\
www.cam.ac.uk}

\author{JUAN MANUEL G\'ORRIZ}

\address{Department of Signal Theory, Networking and Communications, University of Granada, Avenida del Hospicio,\\
Granada, 18071, Spain\\
www.ugr.es}

\author{JOHN SUCKLING}

\address{Department of Psychiatry, University of Cambridge, Herchel Smith Building, Robinson Way,\\
Cambridge, CB20SZ, UK\\
www.cam.ac.uk}

\maketitle

\begin{abstract}
Deep learning models for MRI classification face two recurring problems: they are typically limited by low sample size, and are abstracted by their own complexity (the ``black box problem"). In this paper, we train a convolutional neural network (CNN) with the largest multi-source, functional MRI (fMRI) connectomic dataset ever compiled, consisting of 43,858 datapoints. We apply this model to a cross-sectional comparison of autism (ASD) vs typically developing (TD) controls that has proved difficult to characterise with inferential statistics. To contextualise these findings, we additionally perform classifications of gender and task vs rest. Employing class-balancing to build a training set, we trained 3$\times$300 modified CNNs in an ensemble model to classify fMRI connectivity matrices with overall AUROCs of 0.6774, 0.7680, and 0.9222 for ASD vs TD, gender, and task vs rest, respectively. Additionally, we aim to address the black box problem in this context using two visualization methods. First, class activation maps show which functional connections of the brain our models focus on when performing classification. Second, by analyzing maximal activations of the hidden layers, we were also able to explore how the model organizes a large and mixed-centre dataset, finding that it dedicates specific areas of its hidden layers to processing different covariates of data (depending on the independent variable analyzed), and other areas to mix data from different sources. Our study finds that deep learning models that distinguish ASD from TD controls focus broadly on temporal and cerebellar connections, with a particularly high focus on the right caudate nucleus and paracentral sulcus.
\end{abstract}


\begin{multicols}{2}
\section{Introduction}

The characterization of brain differences in autism spectrum disorder (ASD) is an ongoing challenge. Although the consensus is that there are widespread structural and functional differences, the direction and spatial patterns of differences are not reliably observed and overlap with inter-individual variability in the neurotypical population.

Estimates of grey matter volume with voxel-based morphometry (VBM) have been the most commonly used methodology to assess brain structure, but have resulted in discrepancies amongst meta-analytic findings \cite{Cauda2011,DeRamus2015,Yang2015}, at least a partial explanation for which are the small-sample sizes that are a prevalent feature of the primary literature \cite{Button2013,Nord2017}.

To address variations in data acquisition and processing that make between-study comparisons less powerful, publicly available large-sample datasets are now pivotal to imaging research. The ABIDE multi-centre initiative has made available over 2000 images in two releases, but cross-sectional VBM analyses have failed to observe significant differences \cite{Haar2016,Zhang2018}. Other morphological properties of the cortex may yield greater sensitivity \cite{Khundrakpam2017}, and recent findings using estimates of cortical thickness from the ENIGMA working group suggest a complex pattern of differences relative to neurotypical controls that varies across the lifespan \cite{Rooij2017}. Other databases, such as the National Database for ASD Research (NDAR) act as aggregates of MRI data for different smaller-scale studies, though centre differences complicate conventional analyses on these data as a whole.

ASD has been consistently associated with differences in brain function \cite{Muller2008,Simas2015}. This is often studied in the context of EEG \cite{Ahmadlou2010,Ahmadlou2012,Bhat2014a,Bhat2014b}, for which several studies have been conducted to achieve automated diagnosis \cite{Antoniades2018,Hua2019,Ansari2019,Schaper2019,Acharya2018a,Acharya2018b}, and fMRI. The measurement of correlation, or ``functional connectivity", between time-series of blood oxygenation level dependent (BOLD) endogenous contrast estimated from brain regions during resting wakefulness has been demonstrated as a reproducible measurement on an individual basis \cite{Finn2015}. Functional connectivity (FC) matrices are estimates of the connectivities between all brain regions that can be represented as undirected graphs (connectomes) of nodes (brain regions) and edges (connectivity strengths). They show promise in localising characteristic differences for ASD in resting activity to specific large-scale brain networks \cite{Wang2018}. Whilst there is cautionary evidence using the ABIDE dataset and others \cite{Plitt2015}, it would appear that statistically significant differences in connectivity are generally observable, but like measurements of brain structure, are variable in their presentation. With consistent and localised changes remaining elusive, a number of studies have characterised ASD as exhibiting under-connectivity in certain areas of the brain \cite{Just2004,Cherkassky2006,Kennedy2008,Assaf2010,Jones2010,Weng2010}, while others show evidence of over-connectivity \cite{Cerliani2015,Chien2015,Delmonte2013,Martino2011,Nebel2014a,Nebel2014b}. A recent review \cite{Hull2017} posited that ASD is likely a mix of these traits.

In other fields, computing power and access to large datasets have led to a resurgence in the popularity of NNs as a tool for data classification \cite{LeCun1999,Hinton2006,Krizhevsky2012}. NNs are especially adept at classifying complex data which parametric inferential statistics may fail to fully characterize due to their inherent assumptions. Given that brain function in ASD has been consistently found to be different but in different ways, such a model may be a sensible approach for a comprehensive representation. In parallel, because of their wide applicability in representing complex data such as proteins and social networks, functional connectomes have undergone significant development in terms of global and local topological descriptions. Some recent work has used NNs for processing connectomes, including whole-graph classification, clustering into sub-graphs, and node-wise classification \cite{Bruna2014,Defferrard2016,Hamilton2017,Hechtlinger2017,Kipf2017,Nikolentzos2017}. Previous efforts to classify functional connectivity in ASD on smaller datasets have achieved accuracy rates that have been described as ``modest to conservatively good" \cite{Hull2017}, though these methods have had trouble replicating on different data \cite{Jung2014,Price2014,Iidaka2015}. More recently, the application of convolutional CNNs to ABIDE data has achieved achieved 68\% to 77.3\% classification accuracies. \cite{Subbaraju2017,Brown2018,Hensfield2018,Khosla2018}.

In this article, we leverage publicly available datasets to amass and automatically pre-process a total of 43,838 functional MRIs from nine different collections. To test the application of CNNs to imaging data, we first classify autistic individuals from typically developing (TD) controls. To validate the proposed models, we then classify functional connectivity matrices based on gender and task vs resting state. All classifications were undertaken using a CNN that uniquely encodes multi-layered connectivity matrices, using an original deep learning architecture, partially inspired by Kawahara et al 2017. We opted to use these connectivity matrices as opposed to full fMRI datasets both for memory management purposes (the average fMRI dataset in our collection is 176 MB per file, while the connectivity matrix is just under 500 KB), and for interpretability, as connectivity matrices allow for the direct analysis of both localized areas and connections between areas. Due to the stochastic properties of NNs and set divisions, we use a standard stratified cross-validation strategy, performing each of our tests across 300 independent models using different subsamples and divisions of the total dataset. To incentivise the model to classify based on phenotypic differences rather than centre differences, class-balancing techniques across participant age and collection were used when building the training and test sets, and compared against the fully-inclusive samples.

Key outputs of the CNN are class activation maps \cite{Simonyan2014,Kawahara2017,Selvaraju2017} that highlight areas of the connectome the model preferentially focuses on when performing its classification, and activation maximization \cite{Erhan2009} of a hidden layer that visualizes how the model partitioned the dataset as a whole following classification. We suggest an index to quantify the output of activation maximisation.

\begin{table*}
\tbl{Average populations present for successfully-preprocessed datasets. Some datasets were not labeled with respect to one or more covariates, so counts may not sum to the listed total. \label{tab:table1}}
{\begin{tabular}{@{}lrrrrrrrrrrr@{}}
\toprule
& & & & & Age & & & & Sex & & Disorders \\
Collection & Subjs & Conns & Rest & Task & Min & Max & Mean & Stddev & F & M & Autism \\ \colrule
1000 FC & 764 & 764 & 764 & 0 & 7.88 & 85.00 & 25.76 & 10.18 & 443 & 321 & 0 \\
ABCD & 1319 & 9205 & 4043 & 5162 & 0.42 & 11.08 & 10.08 & 0.65 & 4339 & 4866 & 113 \\
Abide & 193 & 193 & 193 & 0 & 9.00 & 50.00 & 17.81 & 6.69 & 21 & 172 & 94 \\
Abide II & 720 & 761 & 761 & 0 & 5.22 & 55.00 & 14.44 & 7.45 & 174 & 587 & 375 \\
ADNI & 141 & 261 & 261 & 0 & 56.00 & 95.00 & 73.57 & 7.32 & 146 & 115 & 0 \\
BioBank & 11811 & 16970 & 9937 & 7033 & 40.00 & 70.00 & 55.23 & 7.51 & 8752 & 8218 & 8 \\
ICBM & 112 & 381 & 29 & 352 & 19.00 & 74.00 & 43.53 & 14.83 & 188 & 193 & 0 \\
NDAR & 1123 & 8569 & 5952 & 2617 & 0.25 & 55.83 & 18.65 & 7.82 & 4165 & 4404 & 994 \\
Open fMRI & 1443 & 6655 & 1169 & 5486 & 5.89 & 78.00 & 27.22 & 10.40 & 2768 & 3133 & 127 \\
All & 17614 & 43838 & 23109 & 20650 & 0.25 & 95.00 & 33.05 & 20.68 & 20996 & 22009 & 1711 \\
\botrule
\end{tabular}}
\end{table*}

In attempting to classify components of this accumulated dataset, we sought to address the following questions: (1) How effective is our machine learning paradigm at classifying FC in ASD, gender, and resting-state/task? (2) Which areas or networks of the brain do models focus on when undertaking classifications? (3) How does the model partition large datasets during different classification tasks? (4) Can the model effectively classify FCs taken from multiple sources without relying explicitly on centre differences to do so?

\section{Methods}
\subsection{Datasets and preprocessing}

Datasets were acquired from OpenFMRI \cite{Poldrack2013,Poldrack2017}; the Alzheimer’s Disease Neuroimaging Initiative (ADNI); ABIDE\cite{DiMartino2014}; ABIDE II \cite{DiMartino2017}; the Adolescent Brain Cognitive Development (ABCD) Study \cite{Casey2018}; the NIMH Data Archive, including the Research Domain Criteria Database (RDoCdb), the National Database for Clinical Trials (NDCT), and, predominantly, the National Database for Autism Research (NDAR) \cite{Hall2012}; the 1000 Functional Connectomes Project \cite{Dolgin2010}; the International Consortium for Brain Mapping database (ICBM); and the UK Biobank; we refer to each of these sets as \textit{collections}. OpenFMRI, NDAR, ICBM, and the 1000 Functional Connectomes Project are collections that comprise different datasets submitted from unrelated research groups. ADNI, ABIDE, ABIDE II, ABCD, and the UK Biobank are collections that were acquired as part of a larger research initiative.

These data were pre-processed using the fMRI Signal Processing Toolbox (SPT). Following skull-stripping, motion correction was accomplished using SpeedyPP version 2.0, which utilized AFNI tools and wavelet despiking \cite{Patel2014,Patel2016}, with a low-bandpass filter of 0.01Hz, in addition to motion and motion derivative regression. Both functional and structural datasets were non-linearly registered to MNI space and parcellated using the 116-area automated anatomical labeling (AAL) template \cite{Tzourio2002}, which includes subcortical regions. Extracted time series were the means of each AAL region. Each dataset was transformed into $N$ $4\times 116\times 116$ connectivity matrices, using edges weighted by the Pearson correlation of the wavelet coefficients of the pre-processed time-series in each of four frequency scales: 0.1-0.2 Hz, 0.05-0.1 Hz, 0.03-0.05 Hz, and 0.01-0.03 Hz. Wavelet correlation estimates were adjusted from TR rates to equalize the frequency ranges across different collections. Pre-processing was accomplished on a computing cluster over a period of several weeks. Due to the volume of datasets, individualized quality control was not possible. The porportion of datasets failing pre-processing varied by collection.

Across all collections, 70,284 potential datasets were identified of which 67,396 contained suitable functional and structural datasets. Of these, 52,396 succeeded pre-processing to parcellation. However, datasets with regional dropout of greater than 10\% were omitted from the analyses, and redundant datasets across collections were also discarded along with those data with a TR outside of the desired range. In total, 43,838 connectomes from 17,614 unique participants were available for analysis with the NN. Multiple instances of connectomes from the same individuals were used, though they were not shared between the training, validation, and test sets. The numbers of participants, total numbers of datasets used as well as phenotypic distributions, are shown in Table \ref{tab:table1}.

\subsection{Neural Network Model and Training}

\begin{figurehere}
\centering
\includegraphics[width=.95\columnwidth]{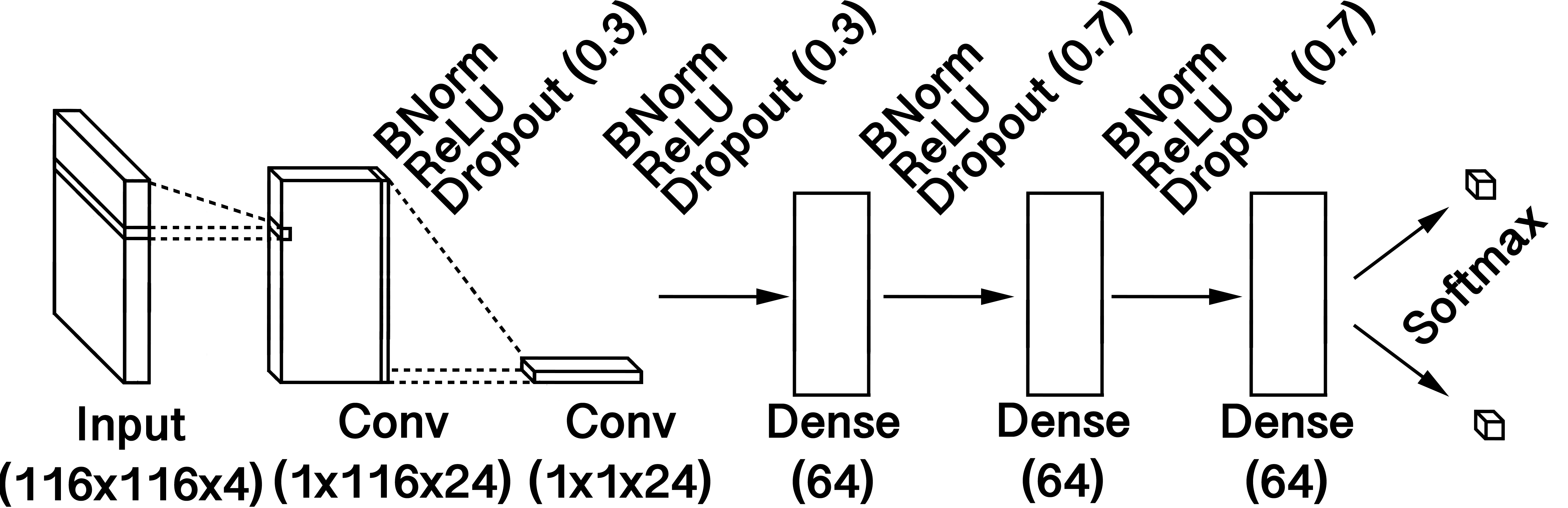}
\caption{The structure of the neural network. These were applied in an ensemble model, so the outputs of 300 independently-trained neural networks were averaged in a cross-validation scheme.}
\label{fig:nn_model}
\end{figurehere}
The data used for training and testing the CNN were $4\times116\times116$ (4 wavelet scales and 116 nodes) symmetric FC (wavelet coefficient correlation) matrices, with values linearly scaled from [-1.1] to [0,1] for easier use in a NN.

To classify the data, we employed a CNN with vertical convolutional filters on the first layer followed by horizontal convolutional filters on the second layer, effectively reducing the matrices to single values to allow the network to train on connectivity matrices (Figure \ref{fig:nn_model}). This approach was partially inspired by the cross-shaped filters described in Kawahara et al 2017, though previous tests with that architecture resulted in a number of failed models with no apparent increase in accuracy over the simpler architecture proposed here. We implemented this architecture using Keras \cite{Chollet2015}, a popular machine learning library, leveraging the advantages of supporting software libraries. Additionally, this implementation includes multiple channels in the inputs, as opposed to single-input connectivity matrices.

The CNN was constructed with: 24 edge-to-node vertical convolutional filters; 24 node-to-graph horizontal convolutional filters; 3 fully-connected layers, each with 64 nodes; and a final softmax layer. Separating each layer were batch normalization, rectified linear unit (ReLU), and dropout layers, with the dropout being 0.3 in the convolutional layers and 0.7 in the dense layers. The layer structures and ordering followed the advice offered in \refcite{Ioffe2015}. Specifications are shown in Figure \ref{fig:nn_model}. No pooling layers were used, and all strides were of length 1. The model was trained using an Adam optimizer with batch sizes of 64. Otherwise, Keras defaults were used. Models were trained for 200 epochs, and the epoch with the highest validation accuracy was selected.

To obtain a reliable average, we trained 300 models independently for each classification, which were then combined in an ensemble model. In each training instance, a subset of the total available data was taken. A holdout test and validation set were not used \cite{Kohavi1995}, but instead a division of the data was performed for each model in a stratified cross-validation schema, subject to the rules detailed below.

\subsection{Set division}

Data were divided into three sets: a training set, comprising of two-thirds of the data and used to train the model; a validation set, comprising of one-sixth of the data and used to select the epoch at which training stopped; and a test set, used to assess the trained classifier performance, comprising of one-sixth of the data. The approximate total number of images used by each model was 10,000 for the gender and resting-state classification, and 4000 (limited by sample size) for the ASD classification.
For all classifications, balancing was used such that each class comprised approximately half of the datasets. To account for covariates, classes were additionally balanced such that the distributions of different collections and ages were equal between classes. For collection balancing, equal numbers of datasets were used from each collections. For continuous age values, distributions of age between classes were made to fail a Mann-Whitney U-test, with $p>0.05$. We used standard stratified cross-validation rather than a holdout division across the 300 runs.

Because of the collection balancing procedure, many data were excluded from certain classification tasks; for instance, as BioBank only included eight subjects with ASD. Due to the class balancing, set divisions were not precise in each instance.

\subsection{Test set evaluation}

\subsubsection{Inter-data classification} 

Following the training of the models, the accuracy and the area under the receiver operating characteristic curve (AUROC) were calculated as measures of machine learning performance on the test set. This was to determine if one group in the classification outperformed the other in training leading to a biasing of the overall accuracy.

\subsubsection{Activation Maximization} 

Activation maximization \cite{Erhan2009} is a technique to determine the maximally activated hidden units in response to the test set of the CNN layers following training. Activation maximization was applied to the $116\times24$ second layer of our network (Figure \ref{fig:nn_model}) as this convolutional layer acts as a bottleneck, and is thus easier to interpret and visualize. This layer is naturally stratified by 24 \textit{filters}, each with 116 nodes (brain regions). To offset the influence of spurious maximizations, we opted to record the 10 datasets that maximally activated each hidden unit, obtaining their mode with respect to collection, gender, and whether it was task/rest; for example, if six connectomes that maximally activated a unit were from Collection A and four were from Collection B, Collection A would be recorded as maximally activating that hidden unit.

For each covariate, this method yields a $116\times24$ array of values for each of the $3\times300$ models. We opted to measure the stratification of the different convolutional filters in our models by measuring whether it was maximally activated primarily by one source of data, or whether it was activated by a mixed population. With this in mind, we calculated for each layer a diversity coefficient, which is 0 if the layer is only maximally activated by one class of data and 1 if it is maximized proportional to the population maximized. Given $K$ possible classes, $F_k, k \in K$ indicating the percentage of each class in a given filter, and $T_k, k \in K$ indicating the percentage of each class across all filters, we calculated the diversity coefficient for each filter as:
\begin{equation}
\resizebox{0.4 \textwidth}{!}{
$ D_i = \frac{\tan^{-1}{\Bigg(
         \ln{\frac{1 - \sqrt{\sum_{k=1}^{K}{F_k^2}}}{\sqrt{\sum_{k=1}^{K}{\frac{(F_k - T_k)^2}{2}}}}}
        \Bigg)} + \frac{\pi}{2}}{\pi}
$
}
\end{equation}

Briefly, the justification for this equation is that the summation $\sum_{k=1}^{K}{\frac{(F_k - T_k)^2}{2}}$ equals 0 if the distribution of the filter's population is equal to the population of the whole layer; that is, the distribution is ideally diverse, and this pulls the logarithm towards $-\infty$, which in turn pulls the inverse tangent function to $\frac{\pi}{2}$. Conversely, $1 - \sqrt{\sum_{k=1}^{K}{F_k^2}}$ tends towards 0 if the individual layer is only composed of a single class, pulling the inverse tangent towards -$\frac{\pi}{2}$. The diversity coefficient is normalized to be between 0 and 1. Its value is indeterminate if only a single class is present globally.

This equation is a more complex version of other diversity coefficients, such as the Herfindahl-Hirschman or Simpson diversity indices. However, the proposed index better accounts for overall populations in the hidden layer activations and thus makes it easier to compare across different classification tasks and independent variables. While the Herfindahl-Hirschman or Simpson indices both approach their maxima when the measured population is completely homogenous, their lower extrema varies depending on the number of distinct populations present. This is problematic in comparing across indices, because the number of populations varies depending on the application, and assumes that the expected (i.e., most diverse) distribution occurs when different populations are perfectly proportional. The proposed index defines the most diverse population as that which has distributions proportional to the overall population, at which point the index is zero.

In practice, low diversity coefficients indicate that the ensemble models stratified data by the covariate. This allows us to measure the degree to which individual covariates (such as collection) were taken into account by the CNNs. We found the diversity coefficient of each of the 24 filters of our hidden, $116\times24$, convolutional layers, then sorted these values to show which filters were primarily activated by a few covariates and which were activated maximally by many covariates.

\begin{tablehere}
\tbl{The ensemble and averaged AUROCS and accuracies for 300 models.\label{tab:table2}}
{\begin{tabular}{@{}llll@{}}
\toprule
 & Autism & Gender & Rest v Task \\ \colrule
Ensemble AUROC & 0.6774 & 0.7680 & 0.9222 \\
Ensemble Acc. & 67.0253\% & 69.7063\% & 85.1996\% \\
Average AUROC & 0.6133 & 0.6858 & 0.9231 \\
Average Acc. & 57.1150\% & 63.3398\% & 84.3153\% \\

\botrule
\end{tabular}}
\end{tablehere}

\subsubsection{Class Activation Maps} 

We used class activation maps (CAMs)\cite{Simonyan2014,Kawahara2017,Selvaraju2017} and a previous Keras implementation \cite{raghakotkerasvis} to display parts of the connectivity matrix the CNN emphasised in its classification of the test sets. CAMs operate by taking the derivative of the CNN classification function (approximated as a first-order Taylor expansion, estimated via back-propagation) with respect to an input matrix, with the output being the same dimensions as the input \cite{Simonyan2014,Selvaraju2017}. While class activation maps were originally proposed in \cite{Simonyan2014}, they were improved to the commonly-used method presented in \cite{Selvaraju2017}, known as `Gradient Class Activation Maps (Grad-CAMs). CAMs are particularly advantageous when applied to connectivity matrices, because unlike typical 2D images, these matrices are spatially static (i.e. each part of the matrix represents the same connection in the brain, across all datasets). Thus, global tendencies of the model can be visualized by averaging many CAMs. CAMs for each connectivity matrix were averaged, maximised across the four wavelet frequency domains, and displayed to show which aspects of the connectome the CNN focused. To simplify the analysis, CAMs were taken with respect to the input data's predicted output, rather than two output classes.

\subsection{Experiments}

We performed the classification on class- and age-balanced datasets that then classified based on gender, task vs rest, and ASD vs TD controls in separate analyses. We then analysed the averaged CAMs with respect to their output prediction. We also recorded the diversity coefficient with respect to gender, collection, and rest v task.
\begin{figurehere}
\centering
\includegraphics[width=0.9\columnwidth]{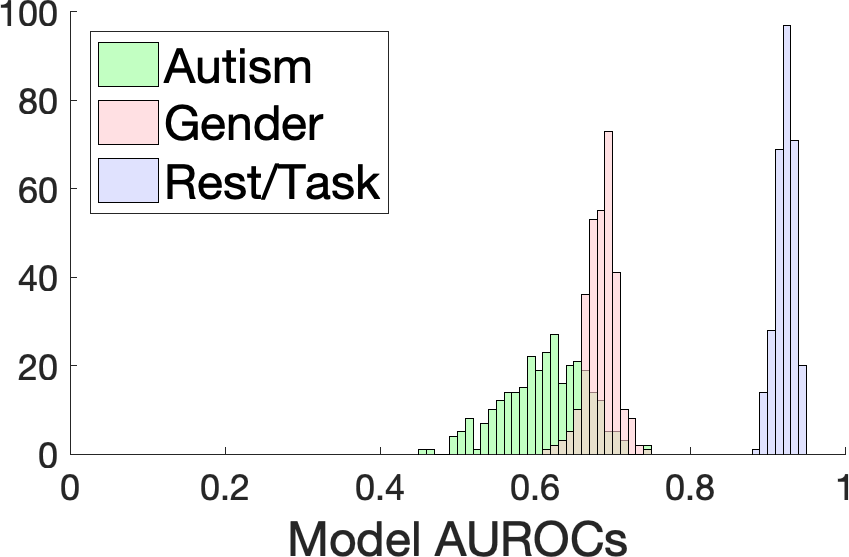}
\caption{Histograms of all AUROCS for 300 independent models, using different, stratified samples of the whole dataset.}
\label{fig:histogram_accs}
\end{figurehere}
\section{Results}

\begin{figure*}
\centering
\raisebox{40ex}{\fontfamily{qhv}{\fontsize{30}{100}{\textbf{A}}}}
\includegraphics[width=0.6\linewidth]{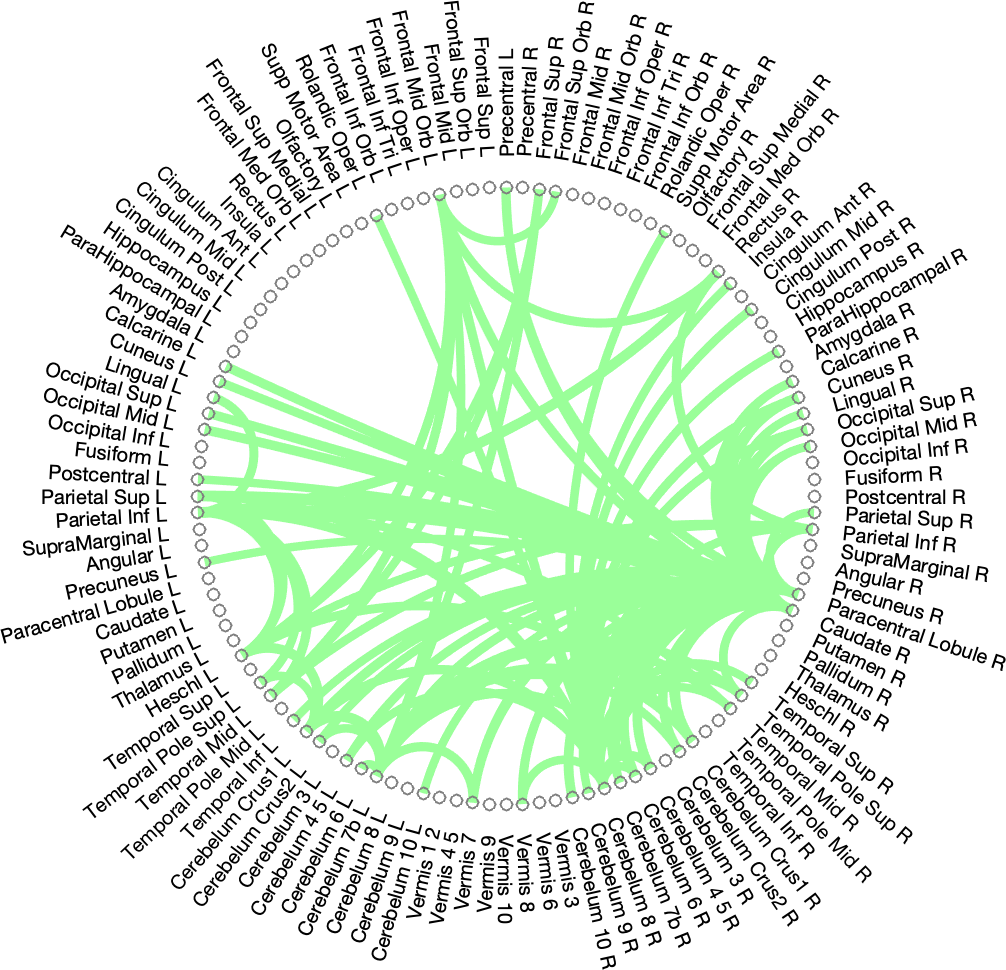}
\raisebox{40ex}{\fontfamily{qhv}{\fontsize{30}{100}{\textbf{B}}}}\includegraphics[width=0.45\linewidth]{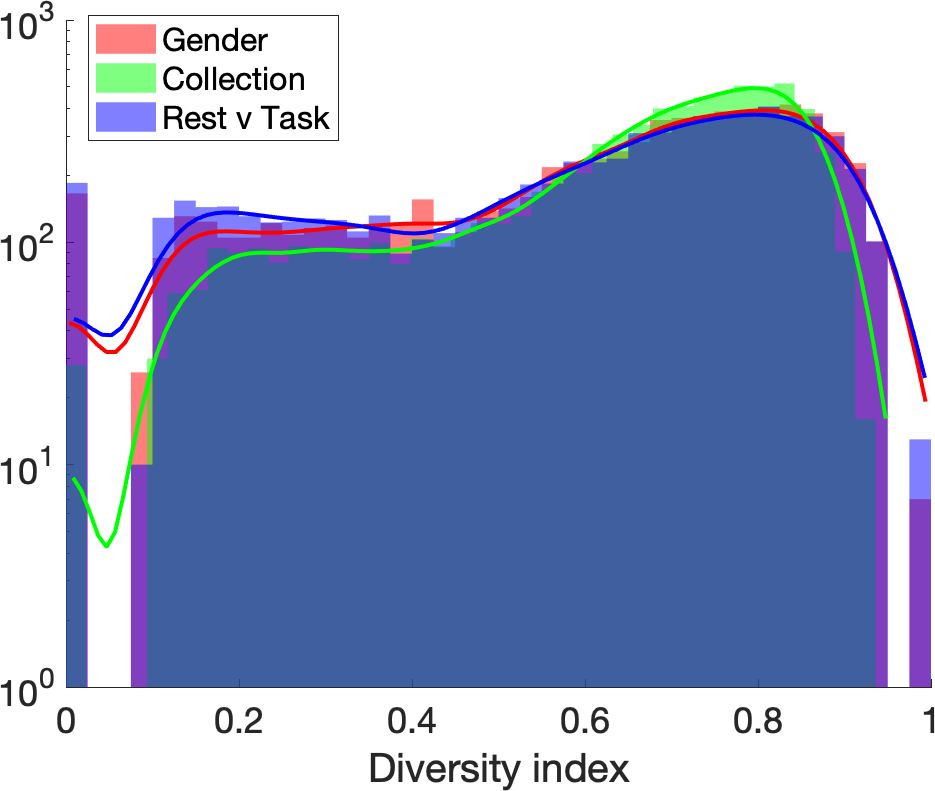}
\raisebox{40ex}{\fontfamily{qhv}{\fontsize{30}{100}{\textbf{C}}}}\includegraphics[width=0.40\linewidth]{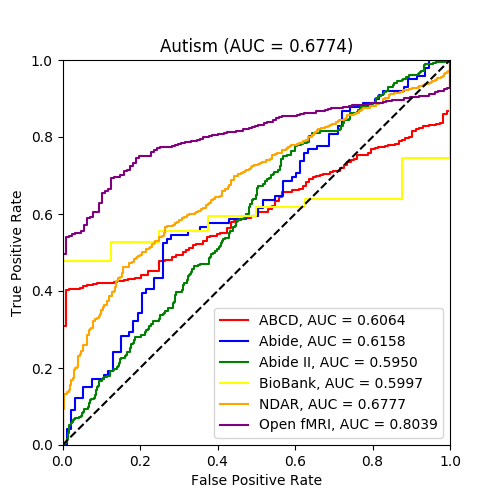}
\caption{Results for autism classification. (A) The 100 strongest connections of the mean class activation maps, with the maximum value taken across wavelet correlations. (B) The distribution of the diversity index of maximal activations across all filters over 300 models, showing how much filters in general were dedicated to particular phenotypes. (C) The overall classification AUROC and the AUROC of individual data collections in the model, showing the overall and relative success of the model.}
\label{fig:asd_or_not_fig}
\end{figure*}

\begin{figure*}
\centering
\raisebox{40ex}{\fontfamily{qhv}{\fontsize{30}{100}{\textbf{A}}}}
\includegraphics[width=0.6\linewidth]{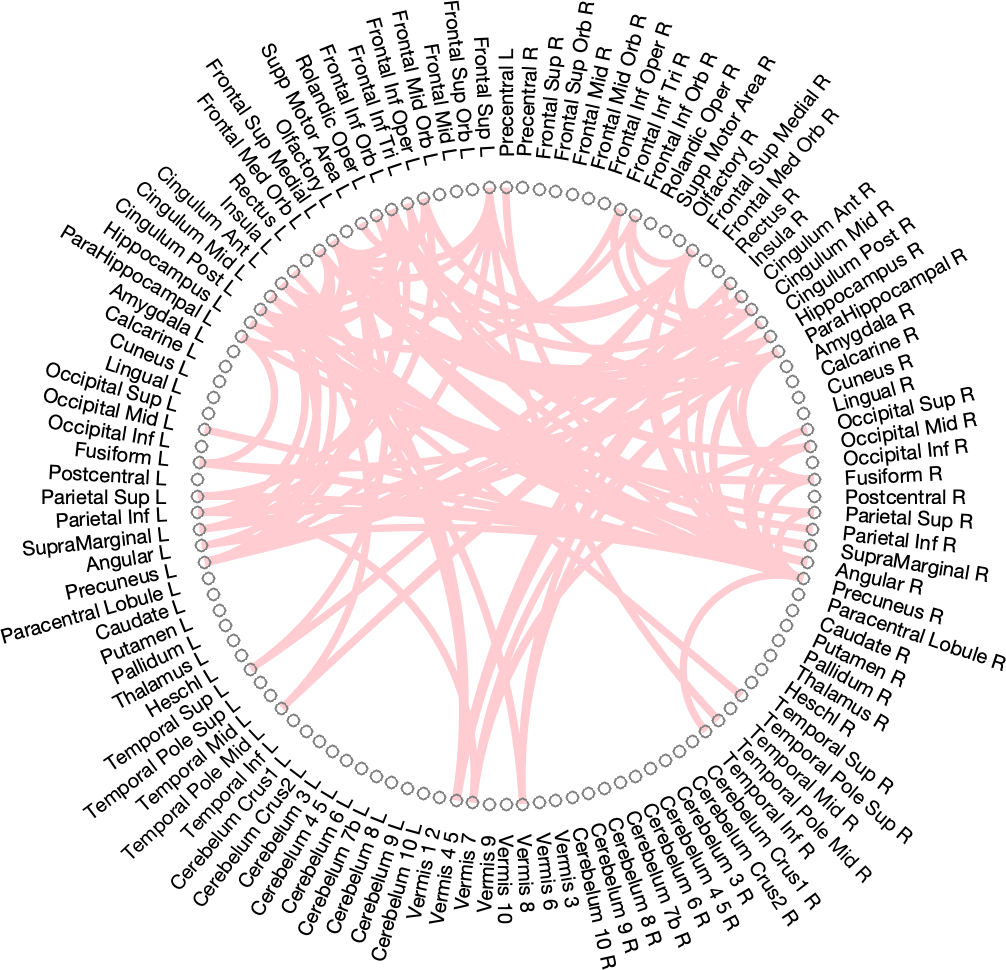}
\raisebox{40ex}{\fontfamily{qhv}{\fontsize{30}{100}{\textbf{B}}}}\includegraphics[width=0.45\linewidth]{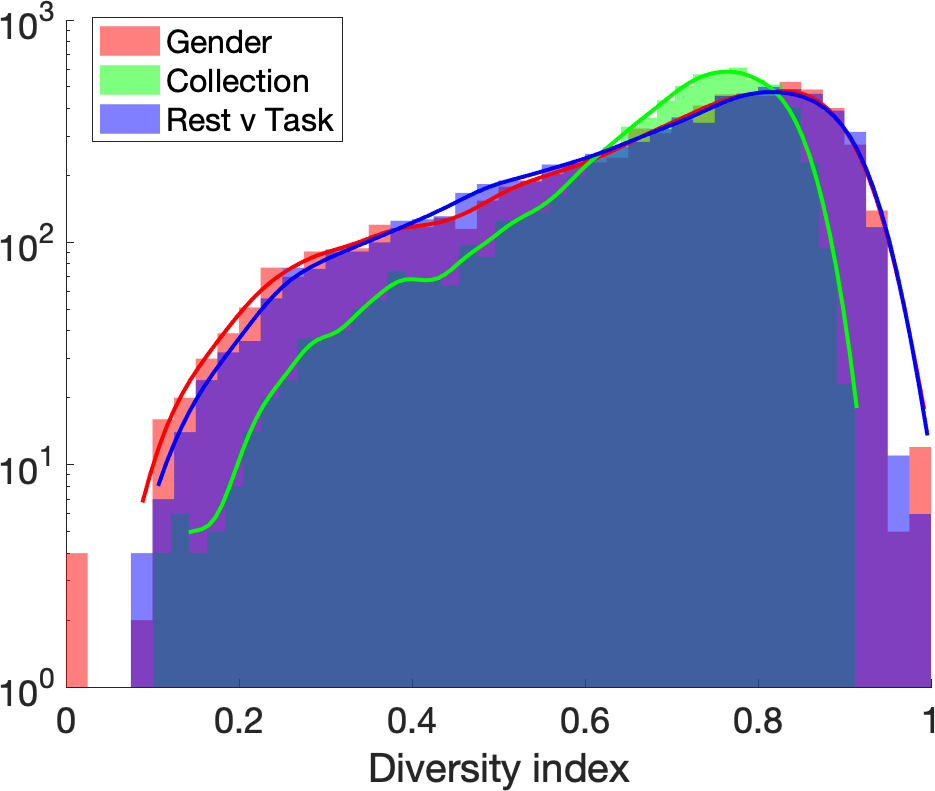}
\raisebox{40ex}{\fontfamily{qhv}{\fontsize{30}{100}{\textbf{C}}}}\includegraphics[width=0.40\linewidth]{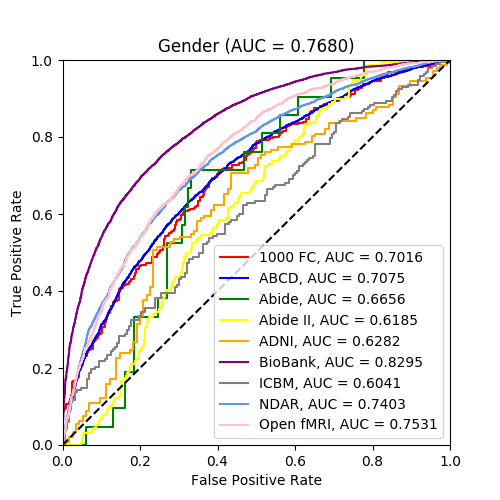}
\caption{Results for gender classification. (A) The 100 strongest connections of the mean class activation maps, with the maximum value taken across wavelet correlations. (B) The distribution of the diversity index of maximal activations across all filters over 300 models, showing how much filters in general were dedicated to particular phenotypes. (C) The overall classification AUROC and the AUROC of individual data collections in the model, showing the overall and relative success of the model.}
\label{fig:gender_fig}
\end{figure*}

Table \ref{tab:table2} shows the accuracies for the 300 models tested.  The AUROCs for the individual models, across all data (Figure \ref{fig:histogram_accs}) were averaged to give 0.6858, 0.9231, and 0.6133 for gender, task vs rest, and ASD vs TD classifications, respectively, while the average accuracies were 63.33\%, 84.31\%, and 57.11\%. In nearly all cases, however, as shown in Table \ref{tab:table2}, the ensemble AUROC and accuracies were substantially higher. The ROC of ensemble models with respect to collections are shown in Figures \ref{fig:asd_or_not_fig}C, \ref{fig:gender_fig}C, and \ref{fig:rest_or_task_fig}C.

The results in Figures \ref{fig:asd_or_not_fig}B, \ref{fig:gender_fig}B, and \ref{fig:rest_or_task_fig}B display the histogram of diversity indices across all models' activation maximisation values. This indicates the tendency of models to use particular filters to sequester data by different covariates, especially if it were attempting to classify by that variable; thus, a diversity index of 0 indicates that all nodes within a particular filter were maximally activated from one or a small number of collections (i.e., BioBank or Open fMRI). The covariates measured are gender, rest/task, and collection site; ASD was not included as a covariate because of the relatively small percentage of ASD data overall.

The diversity index of the activation maximization of the second hidden layer revealed that filters we in many cases sorted into two distinct groups, as shown by peaks on the lower and upper end of histograms in Figures \ref{fig:asd_or_not_fig}B, \ref{fig:gender_fig}B, and \ref{fig:rest_or_task_fig}B: stratified layers (i.e., with a diversity index close to 0), which were wholly maximally activated by one type of dataset, and mixed layers (i.e., with a diversity index close to 1), which integrated data from different sources. While gender and task vs rest each had a proportion of their filters wholly activated by a single collection, the majority of filters were activated by a variety of different collections, indicating the effective synthesis of data from different sources. ASD, however, had a large proportion of data with a diversity index close to zero; this is expected for the gender and resting-state covariates, given that the datasets were mainly from males, but the low diversity indices for collection indicates that ASD classification models sequestered data based on collection, and thus many datasets were considered independently.
\subsection {ASD vs TD Controls}
With class balancing, the ensemble performance for ASD v TD controls across test sets was AUROC=0.6774 (Figure \ref{fig:asd_or_not_fig}). ASD classifications were highly dependent on the collection used, although the final AUROCs were above chance for all collections. Class balancing was particularly necessary for this scheme, as data from autistic individuals comprised less than 10\% overall.

Class activation was strongest for ASD in the limbic system, cerebellum, temporal lobe, and frontal middle orbital lobe, but overwhelmingly emphasised in the right caudate nucleus and paracentral lobule (Figure \ref{fig:asd_or_not_fig}A). Findings of the caudate nucleus are consistent with historical findings in developmental ASD \cite{Qiu2015}, with both aberrant FC frequently associated with that area and the presence of volume differences \cite{Sears1999,McAlonan2002,Brambilla2003,Hollander2005,ODWyer2006,Rojas2006,Turner2006,Qiu2016}.

As stated above, activation maximization saw high stratification with regards to gender and resting-state (Figure \ref{fig:asd_or_not_fig}B). Collection also saw a mix of filters that were both highly stratified and highly diverse, indicating the dual use of convolutional filters. Given the phenotypic differences in our ASD datasets (with ABCD consisting largely of children and ABIDE adolescents, for instance), it is likely that the models considered parts of them independently during classification.

\subsection{Gender}

The ensemble classification of gender yielded 0.7680 AUROC, with comparable AUROCs across different collections (Figure \ref{fig:gender_fig}C).

On average, CAMs in gender classifications showed more differences around areas in the corpus callosum and the frontal lobe (especially the medial left frontal lobe), as well as parietal areas, with very few subcortical differences (Figure \ref{fig:gender_fig}A).

In activation maximization (Figure \ref{fig:gender_fig}B), most of the filters mixed data from different genders and rest/task. A proportion were maximally activated by individual collections, but for the most part, this was mixed as well. Among the three classification tasks in this study, gender integrated the most data from different sources. As gender distributions are likely the most homogenous variable tracked across datasets (with the exception of ABIDE I and II), the stratification with respect to individual collections was appropriately lower than expected when classifying other variables.
\begin{figure*}
\centering
\raisebox{40ex}{\fontfamily{qhv}{\fontsize{30}{100}{\textbf{A}}}}
\includegraphics[width=0.6\linewidth]{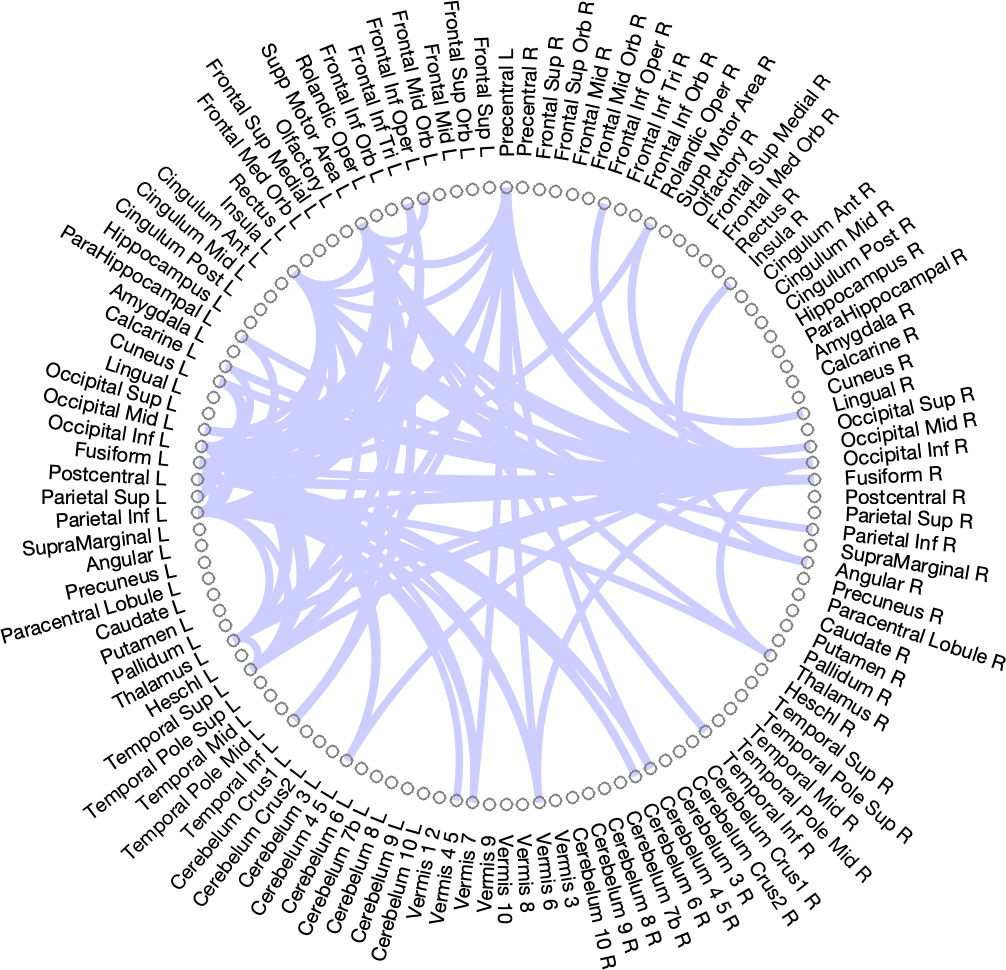}
\raisebox{40ex}{\fontfamily{qhv}{\fontsize{30}{100}{\textbf{B}}}}\includegraphics[width=0.45\linewidth]{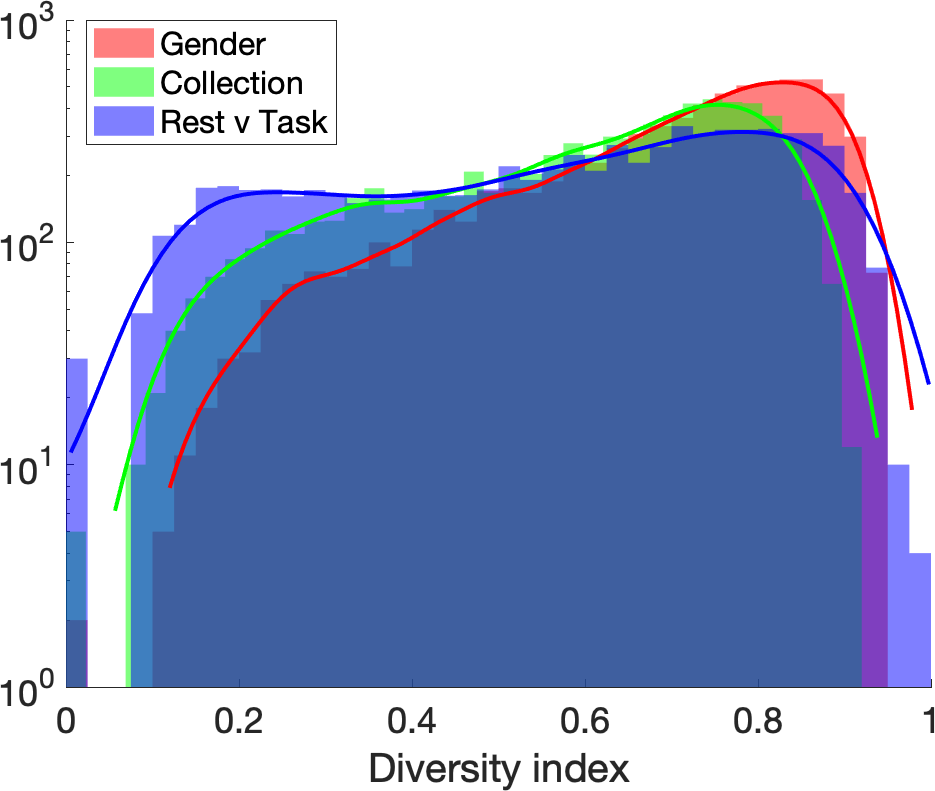}
\raisebox{40ex}{\fontfamily{qhv}{\fontsize{30}{100}{\textbf{C}}}}\includegraphics[width=0.40\linewidth]{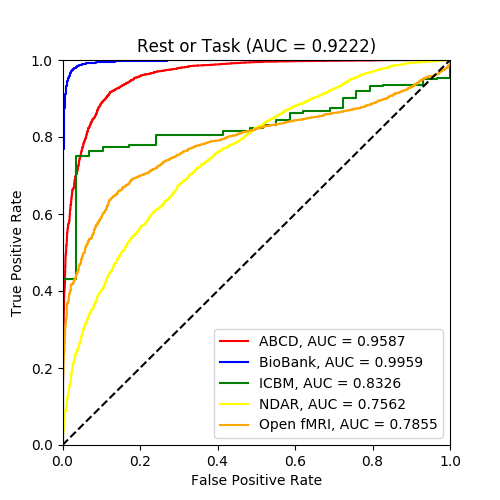}
\caption{Results for resting-state/task classification. (A) The 100 strongest connections of the mean class activation maps, with the maximum value taken across wavelet correlations. (B) The distribution of the diversity index of maximal activations across all filters over 300 models, showing how much filters in general were dedicated to particular phenotypes. (C) The overall classification AUROC and the AUROC of individual data collections in the model, showing the overall and relative success of the model.}
\label{fig:rest_or_task_fig}
\end{figure*}
\subsection {Task vs Rest}
Task v rest classification had an ensemble classification of AUROC=0.9222 (Figure \ref{fig:rest_or_task_fig}C), by far the highest of any classification task. BioBank rest/task classification had nearly perfect classification, while other collections that contributed substantial amounts to both resting-state and task participants, that is, NDAR, ABCD, and Open fMRI, had comparable performance. The CAM focused on the default mode network, largely in the left hemisphere, and its connection to the right frontal medial orbital area. The highly emphasised areas include the supplementary motor area, the left parietal lobe, the bilateral middle and inferior occipital lobe, the left precentral gyrus, and the bilateral thalamus representing the wide range of areas activated in task fMRI.

In activation maximization, stratification was found with respect to task (the target covariate), somewhat on collection, and very little with respect to gender. A degree of collection stratification may be expected due to the different tasks found in different collections; for instance, BioBank consisted almost entirely of an emotional faces recognition task, while Open fMRI contains a medley of different tasks.

\section{Discussion}

This work describes how large and diverse imaging data might be analyzed by deep learning models, encouraging the aggregation of publicly available collections. Data were partitioned based on clear and logical features of the images and, even with imperfect classification accuracies, deep learning models were capable of recognizing complex patterns in large datasets, many consistent with previous work.

The neuroscientific objective of this study was to use available imaging data with deep learning to describe the pattern of functional brain changes that distinguishes ASD from TD. With the absence of any gold standard in this cross-sectional comparison,we also undertook classifications of gender and rest v task, which have more secure, robust findings in the extant literature to confirm the veracity of the developed methods.

We used CAMs \cite{Simonyan2014,Selvaraju2017} to identify connections and areas that had a pronounced influence on the classifications by the model. This method has previously been used in deep learning on functional connectivity \cite{Kawahara2017,Khosla2018} as an effective way of dissecting NNs. However, a caveat to this is that CAMs, while indicators of areas of importance in the data, may not give a complete depiction of its distinguishing features. Without further tests, CAMs cannot indicate whether a particular set of edges is over-connected or under-connected, or whether the areas of high class activation are independent or components of a more complex pattern.

When classifying gender, the model was influenced by diffuse areas connected to the frontal lobe (Figure \ref{fig:gender_fig}A). This is consistent with previous findings in gender comparisons of functional imaging, which did not find differences in brain activity in specific areas, but rather differences in local FC over large areas of the cortex \cite{Tomasi2013}.

Task vs rest FC classifications, as expected, identified the major components of the well-known default mode network \cite{Raichle2001} (Figure \ref{fig:rest_or_task_fig}A), a set of bilateral and symmetric regions that is suppressed during exogenous stimulation \cite{Greicius2003}, as well as visual processing areas (the occipital lobe) and the supplementary motor area. Together with the comparison of gender, the confirmation of the results with those expected from the extant literature give confidence for accurate classification by the CNN as well as the specificity of the visualization method used.

The paracentral lobule and right caudate nucleus, as well as connections to the cerebellum and vermis, were identified as salient to the comparison of the ASD vs TD (Figure \ref{fig:asd_or_not_fig}). This finding is largely substantiated by previous studies that have found both FC and volume differences between autistic and healthy individuals in the caudate nucleus \cite{Sears1999,McAlonan2002,Brambilla2003,Hollander2005,ODWyer2006,Rojas2006,Turner2006}, though these studies disagree on the exact nature of those differences \cite{Qiu2016}. Much of the literature on functional connectivity in ASD, however, concerns network-wide differences \cite{Hull2017} rather than localized differences captured by the CAMs.

Another of the key methods we used to interrogate the results from our deep learning model was activation maximization. Previously, activation maximization has been used for intuiting the internal configuration of NNs rather than for quantitative interpretation \cite{Erhan2009}, which has never been tried, especially across many different independent models. Many of the filters in our models were wholly activated by datasets from a single group, while others utilised a mixture of datasets. We sought to quantify this effect through a diversity index leading to two general observations: first, across models, a few filters were entirely activated by a single collection (i.e., had a diversity index of 0), though which collection remained inconsistent, and was not apparently proportional to the amount of data contributed by that particular dataset; Second, across models, the diversity index was not normally distributed but often had two peaks, one at the low end of the spectrum (indicating stratification of the filters) and one at the high end (indicating a highly diverse, or close to random, distribution of the filter). In ASD, a disproportionately high number of filters were activated by a single collection, indicating that the NN split data internally more than other classification tasks.

In ASD, model accuracy was lower compared to the highest rates reported in literature \cite{Brown2018,Hensfield2018,Khosla2018}, although this result should be viewed with several caveats. The dataset used in this analysis was larger and more varied than any previously analyzed, consisting of many collections. Direct comparisons of machine learning classification methods are difficult as there are no universally accepted schema to divide collections into training and test sets (unlike standardized competitions in other fields, such as the ImageNet Large Scale Visual Recognition Challenge (ILSVRC) \cite{Russakovsky2015}). Furthermore, our exclusion criteria differed, and, because we opted to use multiple scanning sessions from single subjects during training, we also used follow-up data in ABIDE not employed in previous studies. Class balancing may also have significantly affected the classification accuracy. However, this was necessary to avoid spuriously large accuracies due to the highly skewed ratios of ASD-to-TD individuals. Lastly, preprocessing methods and exclusion criteria are not typically shared across collections, and thus technical and demographic differences in the input data cannot be discounted.

While in this study (and all previous large sample-size studies of ASD classification), the classification percentage of ASD v TD datasets does not approach the standards of clinical diagnosis, but remains pertinent. First, the intention of the models is to encourage further research and analysis in this field. Second, FC data may simply lack discrete, distinguishing signals indicative of ASD, making perfect classification impossible, in which case deep learning ought to be viewed as an advanced statistical model rather than a potential diagnostic tool. Third, ASD is a spectrum and not binary (unlike resting-state/task and, in the vast majority of cases, biological gender), and these labels were applied with varying diagnostic standards. While we are simply using the information available, we recognise that the problem itself may be ill-formed. This is also a potential explanation for the variance in model accuracies seen in Figure \ref{fig:histogram_accs}, compared to the other classification problems addressed. Fourth, due to the influence of confounding factors, high accuracy in machine learning for scientific applications should be viewed with skepticism \cite{Ribeiro2016}; for instance, we used several stringent motion-regression algorithms in preprocessing, which likely mitigated the effects of group differences in motion that has previously been observed between autistic and non-autistic subjects \cite{Cook2013}.

Finally, our deep learning model provides several advantages and unique features. First, it employed multichannel input. Although this has long been the standard in 2D image classification (for instance, RGB images), it has not been utilized before in the classification of connectomes. Theoretically, this provides an advantage since it encodes more information about the underlying time-series. In supplementary tests, multichannel inputs generally increased the accuracy of our model by 2–3\% over single-channel Pearson correlation input, though this was not tested extensively. Second, it used vertical filters to encode matrices. In initial versions of this study, we opted to copy the framework of Kawahara et al 2017, which used cross-shaped filters, although this was found to not increase accuracy over vertical filters and caused the model to sometimes fail. Vertical filters were found to be more compatible with the frameworks of modern deep learning libraries, even though they sacrifice the theoretical advantage of encoding edge-to-edge connections.

In our training scheme, we also found substantial accuracy increases with the use of ``ensemble" models in machine learning (Table \ref{tab:table2}); that is, using many independent NNs to vote on a single datapoint. This idea is not new in machine learning \cite{Opitz1999,Polikar2006,Rokach2010}, but it is notable because the ensemble showed a substantial increase in AUROC and accuracy over the sum of the individual models, and thus in this context it was an effective method of smoothing out unexpected behaviour in models for potential real-world applications. Additionally, it is an effective way to evaluate the performance of a model across the entirety of a dataset. Combined with the attractiveness of evaluating and averaging models independently to reduce variance in class activation, this makes a good case for classifying functional connectomes using many independent models rather than one.

\section{Conclusion}

Our investigation was the first to amass an exceedingly large and diverse collection of fMRI data and then apply big data methods. We opted to present three important classification tasks and focus on the one that is both most interesting and least-understood. With careful class-balancing, we show that deep learning models are capable of good-quality classifications across mixed collections detecting differences in brain networks, and functions of localized structures, or FCs over large areas. CAMs highlighted key spatial elements of the classification, and our results were largely validated by prior findings of specific phenotypic differences. Activation maximisation gave insights into the types of features on which the CNN based its classification. While the deep learning model in its present form should not be viewed as a diagnostic tool, it is an example of the apparatus needed to statistically analyse large and publicly accessible volumes of data.

The classification of ASD, on average, pointed overwhelmingly to two key areas (the right caudate nucleus and the right paracentral lobule), which is consistent with many previous studies of ASD. However, it should be noted that the final AUROC was well below the standard for clinical diagnosis, and the variation of model accuracies across our ensemble was very high, especially in relation to the other two categorical classifications. Thus, the areas observed are unlikely to fully characterise ASD. This variation across our very mixed dataset is related to the difficulties of diagnosing ASD in different contexts, and a binary label applied a spectrum disorder may make for an ill-formed machine learning problem.

The most salient future direction of the present work is to focus on one of the classification problems presented and analyse how class activation maps activate differently for different sorts of data. We can also take advantage of several aspects of the technique not explored in the present work, such as comparing the class activations with respect to different input classes. While this was outside the scope of the present study and would have complicated the analysis significantly, it is one of many possible directions in which to take future endeavours.

\nonumsection{Acknowledgments} \noindent This paper is an extension of our previous work at IWINAC 2019 \cite{Leming2019}. This study used publicly available datasets (acknowledgements below). This research was co-funded by the NIHR Cambridge Biomedical Research Centre and Marmaduke Sheild. This work was partly supported by the MINECO/FEDER under the RTI2018-098913-B100 project. Matthew Leming is supported by a Gates Cambridge Scholarship from the University of Cambridge.

\noindent \textbf{ADNI Acknowledgement} Data used in the preparation of this article were obtained from the Alzheimer's Disease Neuroimaging Initiative (ADNI) database (adni.loni.usc.edu). The ADNI was launched in 2003 as a public-private partnership, led by Principal Investigator Michael W. Weiner,MD. The primary goal of ADNI has been to test whether serial magnetic resonance imaging (MRI), positron emission tomography (PET), other biological markers, and clinical and neuropsychological assessment can be combined to measure the progression of mild cognitive impairment (MCI) and early Alzheimer's disease (AD). For up-to-date information, see www.adni-info.org.

\noindent \textbf{ICBM Acknowledgement} Data collection and sharing for this project was provided by the International Consortium for Brain Mapping (ICBM; Principal Investigator: John Mazziotta, MD, PhD). ICBM funding was provided by the National Institute of Biomedical Imaging and BioEngineering. ICBM data are disseminated by the Laboratory of Neuro Imaging at the University of Southern California.

\noindent \textbf{NDAR Acknowledgement} Data and/or research tools used in the preparation of this manuscript were obtained from the NIH-supported
National Database for ASD Research (NDAR). NDAR is a collaborative informatics system created by the National Institutes of Health to provide a national resource to support and accelerate research in ASD. Dataset identifier(s): [NIMH Data Archive Collection ID(s) or NIMH Data Archive Digital Object Identifier (DOI)]. This manuscript reflects the views of the authors and may not reflect the opinions or views of the NIH or of the Submitters submitting original data to NDAR.

\noindent \textbf{ABCD Acknowledgement} Data used in the preparation of this article were obtained from the Adolescent Brain Cognitive Development (ABCD) Study (https://abcdstudy.org), held in the NIMH Data Archive (NDA). This is a multisite, longitudinal study designed to recruit more than 10,000 children age 9-10 and follow them over 10 years into early adulthood. The ABCD Study is supported by the National Institutes of Health and additional federal partners under award numbers U01DA041022, U01DA041028, U01DA041048, U01DA041089, U01DA041106, U01DA041117, U01DA041120, U01DA041134, U01DA041148, U01DA041156, U01DA041174, U24DA041123, and U24DA041147. A full list of supporters is available at https://abcdstudy.org/federal-partners.html. A listing of participating sites and a complete listing of the study investigators can be found at https://abcdstudy.org/Consortium\_Members.pdf. ABCD consortium investigators designed and implemented the study and/or provided data but did not necessarily participate in analysis or writing of this report. This manuscript reflects the views of the authors and may not reflect the opinions or views of the NIH or ABCD consortium investigators. 

\noindent \textbf{UK Biobank Acknowledgement} This research has been conducted using the UK Biobank Resource [project ID 20904], co-funded by the NIHR Cambridge Biomedical Research Centre and a Marmaduke Sheild grant to Richard A.I. Bethlehem and Varun Warrier. The views expressed are those of the author(s) and not necessarily those of the NHS, the NIHR or the Department of Health and Social Care.

\noindent \textbf{Other database Acknowledgements} We would also like to thank the 1000 Functional Connectomes Project, ABIDE I and II, and Open fMRI.
\bibliographystyle{ws-ijns}
\bibliography{references}

\begin{thebibliography}{10}

\bibitem{Cauda2011}
F.~Cauda, E.~Geda, K.~Sacco, F.~D'Agata, S.~Duca, G.~Geminiani and R.~Keller,
  Grey matter abnormality in autism spectrum disorder: an activation likelihood
  estimation meta-analysis study, {\em J Neurol Neurosurg Psychiatry} {\bf 82}
  (2011)  1304--1313.

\bibitem{DeRamus2015}
T.~DeRamus and R.~Kana, Anatomical likelihood estimation meta-analysis of grey
  and white matter anomalies in autism spectrum disorders author links open
  overlay panel, {\em NeuroImage: Clin.} {\bf 7}  (2015)  525--536.

\bibitem{Yang2015}
J.~Yang and J.~Hofmann, Action observation and imitation in autism spectrum
  disorders: an ale meta-analysis of fmri studies, {\em Brain Imaging and
  Behavior} {\bf 10}  (2015)  960--969.

\bibitem{Button2013}
K.~Button, J.~Ioannidis, C.~Mokrysz, B.~Nosek, J.~Flint, E.~J. Robinson and
  M.~Munafo, Power failure: why small sample size undermines the reliability of
  neuroscience, {\em Nat Rev Neurosci} {\bf 14}  (2013)  365--376.

\bibitem{Nord2017}
C.~Nord, V.~Valton, J.~Wood and J.~Roiser, Power-up: A reanalysis of 'power
  failure' in neuroscience using mixture modeling, {\em J Neurosci} {\bf 37}
  (2017)  8051--8061.

\bibitem{Haar2016}
S.~Haar, S.~Berman, M.~Behrmann and I.~Dinstein, Anatomical abnormalities in
  autism?, {\em Cereb Cortex} {\bf 26}  (2016)  1440--1452.

\bibitem{Zhang2018}
W.~Zhang, W.~Groen, M.~Mennes, C.~Greven, J.~Buitelaar and N.~Rommelse,
  Revisiting subcortical brain volume correlates of autism in the abide
  dataset: effects of age and sex, {\em Psychological Medicine} {\bf 48}
  (2018)  654--668.

\bibitem{Khundrakpam2017}
B.~Khundrakpam, J.~Lewis, P.~Kostopoulos, F.~Carbonell and A.~Evans, Cortical
  thickness abnormalities in autism spectrum disorders through late childhood,
  adolescence, and adulthood: A large-scale mri study, {\em Cereb Cortex} {\bf
  27}  (2017)  1721--1731.

\bibitem{Rooij2017}
D.~van Rooij, E.~Anagnostou, C.~Arango, G.~Auzias, M.~Behrmann, G.~Busatto,
  S.~Calderoni, E.~Daly, C.~Deruelle, A.~Di~Martino {\em et~al.}, Cortical and
  subcortical brain morphometry differences between patients with autism
  spectrum disorder and healthy individuals across the lifespan: Results from
  the enigma asd working group, {\em American Journal of Psychiatry} {\bf 175}
  (2017)  359--369.

\bibitem{Muller2008}
E.~M\"uller, A.~Schuler and G.~Yates, Social challenges and supports from the
  perspective of individuals with asperger syndrome and other autism spectrum
  disabilities, {\em Autism} {\bf 12}  (2008)  173--190.

\bibitem{Simas2015}
T.~Simas, S.~Chattopadhyay, C.~Hagan, P.~Kundu, A.~Patel, R.~Holt, D.~Floris,
  J.~Graham, C.~Ooi, R.~Tait {\em et~al.}, Semi-metric topology of the human
  connectome: Sensitivity and specificity to autism and major depressive
  disorder, {\em PLoS One} {\bf 10}  (2015).

\bibitem{Ahmadlou2010}
M.~Ahmadlou, H.~Adeli and A.~Adeli, Fractality and a wavelet-chaos-neural
  network methodology for eeg-based diagnosis of autistic spectrum disorder,
  {\em Journal of Clinical Neurophysiology} {\bf 27}  (2010)  328--333.

\bibitem{Ahmadlou2012}
M.~Ahmadlou, H.~Adeli and A.~Adeli, Fuzzy synchronization likelihood-wavelet
  methodology for diagnosis of autism spectrum disorder, {\em Journal of
  Neuroscience Methods} {\bf 211}  (2012)  203--209.

\bibitem{Bhat2014a}
S.~Bhat, U.~Acharya, H.~Adeli, G.~Muralidhar~Bairy and A.~Adeli, Automated
  diagnosis of autism: In search of a mathematical marker, {\em Reviews in the
  Neurosciences} {\bf 25}  (2014)  851--861.

\bibitem{Bhat2014b}
S.~Bhat, U.~Acharya, G.~Adeli, Muralidhar~Bairy,  and A.~Adeli, Autism: Cause
  factors, early diagnosis and therapies, {\em Reviews in the Neurosciences}
  {\bf 25}  (2014)  841--850.

\bibitem{Antoniades2018}
A.~Antoniades, L.~Spyrou, D.~Martin-Lopez, A.~Valentin, G.~Alarcon, S.~Sanei
  and C.~Took, Deep neural architectures for mapping scalp to intracranial eeg,
  {\em Intl J Neur Sys} {\bf 28}  (2018).

\bibitem{Hua2019}
C.~Hua, H.~Wang, H.~Wang, S.~Lu, C.~Liu,  and S.~Khalid, A novel method of
  building functional brain network using deep learning algorithm with
  application in proficiency detection, {\em Intl J Neur Sys} {\bf 29}  (2019)
  p. 1850015.

\bibitem{Ansari2019}
A.~Ansari, P.~Cherian, A.~Caicedo, G.~Naulaers, M.~De~Vos and S.~Van~Huffel,
  Neonatal seizure detection using deep convolutional neural networks, {\em
  Intl J Neur Sys} {\bf 29}  (2019) p. 1850011.

\bibitem{Schaper2019}
F.~Schaper, Y.~Zhao, M.~Janssen, G.~Wagner, A.~Colon, D.~Hilkman, E.~Gommer,
  M.~Vlooswijk, G.~Hoogland, L.~Ackermans {\em et~al.}, Single cell recordings
  to target the anterior nucleus of the thalamus in deep brain stimulation for
  patients with refractory epilepsy, {\em Intl J Neur Sys} {\bf 29}  (2019).

\bibitem{Acharya2018a}
U.~Acharya, S.~Oh, Y.~Hagiwara, J.~Tan, H.~Adeli and D.~Subha, Automated
  eeg-based screening of depression using deep convolutional neural network,
  {\em Comp Meth and Progr in Biomed} {\bf 161}  (2018)  103--113.

\bibitem{Acharya2018b}
U.~Acharya, S.~Oh, Y.~Hagiwara, J.~Tan,  and H.~Adeli, Deep convolutional
  neural network for the automated detection of seizure using eeg signals, {\em
  Computers in Biology and Medicine} {\bf 100}  (2018)  270--278.

\bibitem{Finn2015}
E.~Finn, X.~Shen, D.~Scheinost, M.~Rosenberg, J.~Huang, M.~Chun,
  X.~Papademetris and R.~Constable, Functional connectome fingerprinting:
  Identifying individuals using patterns of brain connectivity, {\em Nat
  Neurosci.} {\bf 18}  (2015)  1664--1671.

\bibitem{Wang2018}
W.~Wang, J.~Liu, S.~Shi, T.~Liu, L.~Ma, X.~Ma, J.~Tian, Q.~Gong and M.~Wang,
  Altered resting-state functional activity in patients with autism spectrum
  disorder: A quantitative meta-analysis, {\em Front. Neurol.} {\bf 9}  (2018)
  p. 556.

\bibitem{Plitt2015}
M.~Plitt, K.~Barnes and A.~Martin, Functional connectivity classification of
  autism identifies highly predictive brain features but falls short of
  biomarker standards, {\em NeuroImage Clinical} {\bf 7}  (2015)  359--66.

\bibitem{Just2004}
M.~Just, V.~Cherkassky, T.~Keller and N.~Minshew, Cortical activation and
  synchronization during sentence comprehension in high-functioning autism:
  evidence of underconnectivity, {\em Brain} {\bf 127}  (2004)  1811--21.

\bibitem{Cherkassky2006}
V.~Cherkassky, R.~Kana, T.~Keller and M.~Just, Functional connectivity in a
  baseline resting-state network in autism, {\em Neuroreport} {\bf 17}  (2006)
  1687--90.

\bibitem{Kennedy2008}
D.~Kennedy and E.~Courchesne, The intrinsic functional organization of the
  brain is altered in autism, {\em NeuroImage} {\bf 39}  (2008)  1877--85.

\bibitem{Assaf2010}
M.~Assaf, K.~Jagannathan, V.~Calhoun, L.~Miller, M.~Stevens, R.~Sahl,
  J.~O'Boyle, R.~Schultz and G.~Pearlson, Abnormal functional connectivity of
  default mode sub-networks in autism spectrum disorder patients, {\em
  NeuroImage} {\bf 53}  (2010)  247--256.

\bibitem{Jones2010}
T.~Jones, P.~Bandettini, L.~Kenworthy, L.~Case, S.~Milleville, A.~Martin and
  R.~Birn, Sources of group differences in functional connectivity: an
  investigation applied to autism spectrum disorder, {\em NeuroImage} {\bf 49}
  (2010)  401--414.

\bibitem{Weng2010}
S.~Weng, J.~Wiggins, S.~Peltier, M.~Carrasco, S.~Risi, C.~Lord and C.~Monk,
  Alterations of resting state functional connectivity in the default network
  in adolescents with autism spectrum disorders, {\em Brain Res} {\bf 1313}
  (2010)  202--214.

\bibitem{Cerliani2015}
L.~Cerliani, M.~Mennes, R.~Thomas, A.~Martino, M.~Thioux and C.~Keysers,
  Increased functional connectivity between subcortical and cortical
  resting-state networks in autism spectrum disorder, {\em JAMA Psychiatry}
  {\bf 72}  (2015)  767--777.

\bibitem{Chien2015}
H.~Chien, H.~Lin, M.~Lai, S.~Gau and W.~Tseng, Hyperconnectivity of the right
  posterior temporo-parietal junction predicts social difficulties in boys with
  autism spectrum disorder, {\em Autism Res} {\bf 8}  (2015)  427--441.

\bibitem{Delmonte2013}
S.~Delmonte, L.~O'Gallagher, E.~Hanlon, J.~McGrath and J.~Balsters, Functional
  and structural connectivity of frontostriatal circuitry in autism spectrum
  disorder, {\em Front Hum Neurosci} {\bf 7}  (2013) p. 430.

\bibitem{Martino2011}
A.~Di~Martino, C.~Kelly, R.~Grzadzinski, X.~Zuo, M.~Mennes, M.~Mairena,
  C.~Lord, F.~Castellanos and M.~Milham, Aberrant striatal functional
  connectivity in children with autism, {\em Biol Psychiatry} {\bf 69}  (2011)
  847--856.

\bibitem{Nebel2014a}
M.~Nebel, A.~Eloyan, A.~Barber and S.~Mostofsky, Precentral gyrus functional
  connectivity signatures of autism, {\em Front Syst Neurosci} {\bf 8}  (2014)
  p.~80.

\bibitem{Nebel2014b}
M.~Nebel, S.~Joel, J.~Muschelli, A.~Barber, B.~Caffo, J.~Pekar and
  S.~Mostofsky, Disruption of functional organization within the primary motor
  cortex in children with autism, {\em Hum Brain Mapp} {\bf 35}  (2014)
  567--580.

\bibitem{Hull2017}
J.~Hull, L.~Dokovna, Z.~Jacokes, C.~Torgerson, A.~Irimia and J.~van Horn,
  Resting-state functional connectivity in autism spectrum disorders: A review,
  {\em Front Psychiatry} {\bf 7}  (2017).

\bibitem{LeCun1999}
Y.~LeCun, P.~Haffner, L.~Bottou and Y.~Bengio, Object recognition with
  gradient-based learning, {\em Lecture Notes in Computer Science} {\bf 1681}
  (1999)  319--345.

\bibitem{Hinton2006}
G.~Hinton, S.~Osindero and Y.-H. Teh, A fast learning algorithm for deep belief
  nets, {\em Neural Computation} {\bf 18}  (2006)  1527--1554.

\bibitem{Krizhevsky2012}
A.~Krizhevsky, I.~Sutskever and G.~Hinton, Imagenet classification with deep
  convolutional neural networks, {\em Advances in Neural Information Processing
  Systems}   (2012).

\bibitem{Bruna2014}
J.~Bruna, W.~Zaremba, , A.~Szlam and Y.~LeCun, Spectral networks and locally
  connected networks on graphs, {\em ICLR}   (2014).

\bibitem{Defferrard2016}
M.~Defferrard, P.~Bresson and X.~Vandergheynst, Convolutional neural networks
  on graphs with fast localized spectral filtering, {\em NIPS}   (2016)
  3844--3852.

\bibitem{Hamilton2017}
W.~Hamilton, R.~Ying and J.~Leskovec, Representation learning on graphs:
  Methods and applications, {\em Bulletin of the IEEE Computer Society
  Technical Committee on Data Engineering}   (2017).

\bibitem{Hechtlinger2017}
Y.~Hechtlinger, P.~Chakravarti and J.~Qin, A generalization of convolutional
  neural networks to graph-structured data, {\em arXiv}   (2017).

\bibitem{Kipf2017}
T.~Kipf and M.~Welling, Semi-supervised classification with graph convolutional
  neural networks, {\em ICLR 2017}   (2017).

\bibitem{Nikolentzos2017}
G.~Nikolentzos, P.~Meladianos, A.~Tixier, K.~Skianis and M.~Vazirgiannis,
  Kernel graph convolutional neural networks, {\em ICANN 2018}   (2017).

\bibitem{Jung2014}
M.~Jung, H.~Kosaka, D.~Saito, M.~Ishitobi, T.~Morita, K.~Inohara, M.~Asano,
  S.~Arai, T.~Munesue, A.~Tomoda {\em et~al.}, Default mode network in young
  male adults with autism spectrum disorder: relationship with autism spectrum
  traits, {\em Mol Autism} {\bf 5}  (2014) p.~35.

\bibitem{Price2014}
T.~Price, C.~Wee, W.~Gao and D.~Shen, Multiple-network classification of
  childhood autism using functional connectivity dynamics, {\em Med Image
  Comput Comput Assist Interv} {\bf 17}  (2014)  177--184.

\bibitem{Iidaka2015}
T.~Iidaka, Resting state functional magnetic resonance imaging and neural
  network classified autism and control, {\em Cortex} {\bf 63}  (2015)  55--67.

\bibitem{Subbaraju2017}
V.~Subbaraju, M.~Suresh, S.~Sundaram and S.~Narasimhan, Identifying differences
  in brain activities and an accurate detection of autism spectrum disorder
  using resting state functional-magnetic resonance imaging: A spatial
  filtering approach, {\em Med Image Anal} {\bf 35}  (2017)  375--389.

\bibitem{Brown2018}
C.~Brown, J.~Kawahara and G.~Hamarneh, Connectome priors in deep neural
  networks to predict autism, {\em ISBI 2018}   (2018).

\bibitem{Hensfield2018}
A.~Heinsfeld, A.~Franco, R.~Craddock, A.~Buchweitz and F.~Meneguzzia,
  Identification of autism spectrum disorder using deep learning and the abide
  dataset, {\em NeuroImage: Clinical} {\bf 17}  (2018)  16--23.

\bibitem{Khosla2018}
M.~Khosla, K.~Jamison, A.~Kuceyeski and M.~Sabuncu, 3d convolutional neural
  networks for classification of functional connectomes, {\em MICCAI 2018}
  (2018).

\bibitem{Simonyan2014}
K.~Simonyan, A.~Vedaldi and A.~Zisserman, Deep inside convolutional networks:
  Visualising image classification models and saliency maps, {\em Workshop at
  International Conference on Learning Representations\/}, 2014.

\bibitem{Kawahara2017}
J.~Kawahara, C.~Brown, S.~Miller, B.~Booth, V.~Chau, R.~Grunau, J.~Zwicker and
  G.~Hamarneh, Brainnetcnn: Convolutional neural networks for brain networks;
  towards predicting neurodevelopment, {\em NeuroImage} {\bf 146}  (2017)
  1038--1049.

\bibitem{Selvaraju2017}
R.~Selvaraju, M.~Cogswell, A.~Das, R.~Vedantam, D.~Parikh and D.~Batra,
  Grad-cam: Visual explanations from deep networks via gradient-based
  localization, {\em 2017 IEEE International Conference on Computer Vision
  (ICCV)}   (2017).

\bibitem{Erhan2009}
D.~Erhan, Y.~Bengio, A.~Courville,  and P.~Vincent, Visualizing higher-layer
  features of a deep network, technical report 1341, University of Montreal
  (2009).

\bibitem{Poldrack2013}
R.~Poldrack, D.~Barch, J.~Mitchell, T.~Wager, A.~Wagner, J.~Devlin, C.~Cumba,
  O.~Koyejo and M.~Milham, Toward open sharing of task-based fmri data: the
  openfmri project, {\em Front Neuroinform} {\bf 7}  (2013).

\bibitem{Poldrack2017}
R.~Poldrack and K.~Gorgolewski, Openfmri: Open sharing of task fmri data, {\em
  NeuroImage} {\bf 144}  (2017)  259--261.

\bibitem{DiMartino2014}
A.~Di~Martino, C.~Yan, Q.~Li, E.~Denio, F.~Castellanos, K.~Alaerts,
  J.~Anderson, M.~Assaf, S.~Bookheimer, M.~Dapretto {\em et~al.}, The autism
  brain imaging data exchange: towards a large-scale evaluation of the
  intrinsic brain architecture in autism, {\em Mol Psychiatry} {\bf 19}  (2014)
   659--67.

\bibitem{DiMartino2017}
A.~Di~Martino, D.~O'Connor, B.~Chen, K.~Alaerts, J.~Anderson, M.~Assaf,
  J.~Balsters, L.~Baxter, A.~Beggiato, S.~Bernaerts {\em et~al.}, Enhancing
  studies of the connectome in autism using the autism brain imaging data
  exchange ii, {\em Sci Data} {\bf 4}  (2017) p. 170010.

\bibitem{Casey2018}
B.~Casey and A.~Dale, The adolescent brain cognitive development (abcd) study:
  Imaging acquisition across 21 sites, {\em Dev Cog Neuro} {\bf 32}  (2018)
  43--54.

\bibitem{Hall2012}
D.~Hall, M.~Huerta, M.~McAuliffe and G.~Farber, Sharing heterogeneous data: The
  national database for autism research, {\em Neuroinf.} {\bf 10}  (2012)
  331--339.

\bibitem{Dolgin2010}
E.~Dolgin, This is your brain online: the functional connectomes project, {\em
  Nat. Med.} {\bf 16}  (2010) p. 351.

\bibitem{Patel2014}
A.~Patel, P.~Kundu, M.~Rubinov, P.~Jones, P.~Vertes, K.~Ersche, J.~Suckling and
  E.~Bullmore, A wavelet method for modeling and despiking motion artifacts
  from resting-state fmri time series, {\em NeuroImage} {\bf 95}  (2014)
  287--304.

\bibitem{Patel2016}
A.~Patel and E.~Bullmore, A wavelet-based estimator of the degrees of freedom
  in denoised fmri time series for probabilistic testing of functional
  connectivity and brain graphs, {\em NeuroImage} {\bf 142}  (2016)  14--26.

\bibitem{Tzourio2002}
N.~Tzourio-Mazoyer, B.~Landeau, D.~Papathanassiou, F.~Crivello, O.~Etard,
  N.~Delcroix, B.~Mazoyer and M.~Joliot, Automated anatomical labeling of
  activations in spm using a macroscopic anatomical parcellation of the mni mri
  single-subject brain, {\em NeuroImage} {\bf 15}  (2002)  273--289.

\bibitem{Chollet2015}
F.~Chollet, keras https://github.com/fchollet/keras,  (2015).

\bibitem{Ioffe2015}
S.~Ioffe and C.~Szegedy, Batch normalization: Accelerating deep network
  training by reducing internal covariate shift, {\em arXiv}   (2015).

\bibitem{Kohavi1995}
R.~Kohavi, A study of cross-validation and bootstrap for accuracy estimation
  and model selection, {\em Intelligence - Volume 2, IJCAI’95, Morgan
  Kaufmann Publishers Inc., San Francisco, C}   (1995)  1137--1143.

\bibitem{raghakotkerasvis}
R.~Kotikalapudi and contributors, keras-vis
  https://github.com/raghakot/keras-vis,  (2017).

\bibitem{Qiu2015}
A.~Qiu, T.~Anh, Y.~Li, H.~Chen, A.~Rifkin-Graboi, B.~Broekman, K.~Kwek, S.~Saw,
  Y.~Chong, P.~Gluckman {\em et~al.}, Prenatal maternal depression alters
  amygdala functional connectivity in 6-month-old infants, {\em Transl
  Psychiatry} {\bf 5}  (2015) p. e508.

\bibitem{Sears1999}
L.~Sears, C.~Vest, S.~Mohamed, J.~Bailey, B.~Ranson and J.~Piven, An mri study
  of the basal ganglia in autism, {\em Prog Neuropsychopharmacol Biol
  Psychiatry} {\bf 23}  (1999)  613--624.

\bibitem{McAlonan2002}
G.~McAlonan, E.~Daly, V.~Kumari, H.~Critchley, T.~van Amelsvoort, J.~Suckling,
  A.~Simmons, T.~Sigmundsson, K.~Greenwood, A.~Russell {\em et~al.}, Brain
  anatomy and sensorimotor gating in asperger's syndrome, {\em Brain} {\bf 125}
   (2002)  1594--1606.

\bibitem{Brambilla2003}
P.~Brambilla, A.~Hardan, S.~di~Nemi, J.~Perez, J.~Soares and F.~Barale, Brain
  anatomy and development in autism: review of structural mri studies, {\em
  Brain Res Bull} {\bf 61}  (2003)  557--569.

\bibitem{Hollander2005}
E.~Hollander, E.~Anagnostou, W.~Chaplin, K.~Esposito, M.~Haznedar, E.~Licalzi,
  S.~Wasserman, L.~Soorya and M.~Buchsbaum, Striatal volume on magnetic
  resonance imaging and repetitive behaviors in autism, {\em Biol Psychiatry}
  {\bf 58}  (2005)  226--232.

\bibitem{ODWyer2006}
L.~O'Dwyer, C.~Tanner, E.~van Dongen, C.~Greven, J.~Bralten, M.~Zwiers,
  B.~Franke, D.~Heslenfeld, J.~Oosterlaan, P.~Hoekstra {\em et~al.}, Decreased
  left caudate volume is associated with increased severity of autistic-like
  symptoms in a cohort of adhd patients and their unaffected siblings, {\em
  PLoS One} {\bf 11}  (2016) p. e0165620.

\bibitem{Rojas2006}
D.~Rojas, E.~Peterson, E.~Winterrowd, M.~Reite, S.~Rogers and J.~Tregellas,
  Regional gray matter volumetric changes in autism associated with social and
  repetitive behavior symptoms, {\em BMC Psych} {\bf 6}  (2006) p.~56.

\bibitem{Turner2006}
K.~Turner, L.~Frost, D.~Linsenbardt, J.~McIlroy and R.~M\:uller, Atypically
  diffuse functional connectivity between caudate nuclei and cerebral cortex in
  autism, {\em Behav Brain Funct} {\bf 2}  (2006).

\bibitem{Qiu2016}
T.~Qiu, C.~Chang, Y.~Li, L.~Qian, C.~Xiao, T.~Xiao, X.~Xiao, Y.~Xiao, K.~Chu
  {\em et~al.}, Two years changes in the development of caudate nucleus are
  involved in restricted repetitive behaviors in 2--5-year-old children with
  autism spectrum disorder, {\em Developmental Cognitive Neuroscience} {\bf 19}
   (2016)  137--143.

\bibitem{Tomasi2013}
D.~Tomasi and N.~Volkow, Gender differences in brain functional connectivity
  density, {\em Hum Brain Mapp.} {\bf 33}  (2013)  849--860.

\bibitem{Raichle2001}
M.~Raichle, A.~MacLeod, A.~Snyder, W.~Powers, D.~Gusnard and G.~Shulman, A
  default mode of brain function, {\em PNAS} {\bf 98}  (2001)  676--682.

\bibitem{Greicius2003}
M.~Greicius, B.~Krasnow, A.~Reiss and V.~Menon, Functional connectivity in the
  resting brain: A network analysis of the default mode hypothesis, {\em PNAS}
  {\bf 100}  (2003)  253--258.

\bibitem{Russakovsky2015}
O.~Russakovsky, J.~Deng, H.~Su, J.~Krause, S.~Satheesh, S.~Ma, Z.~Huang,
  A.~Karpathy, A.~Khosla {\em et~al.}, Imagenet large scale visual recognition
  challenge, {\em International Journal of Computer Vision} {\bf 115}  (2015)
  211--252.

\bibitem{Ribeiro2016}
M.~Ribeiro, S.~Singh and C.~Guestrin, ``why should i trust you?" explaining the
  predictions of any classifier, {\em Knowledge Discovery and Data Mining
  (KDD)}   (2016).

\bibitem{Cook2013}
J.~Cook, S.~Blakemore and C.~Press, Atypical basic movement kinematics in
  autism spectrum conditions, {\em Brain} {\bf 136}  (2013)  2816--2824.

\bibitem{Opitz1999}
D.~Opitz and R.~Maclin, Popular ensemble methods: An empirical study, {\em
  Journal of Artificial Intelligence Research} {\bf 11}  (1999)  169--198.

\bibitem{Polikar2006}
R.~Polikar, Ensemble based systems in decision making, {\em IEEE Circuits and
  Systems Magazine} {\bf 6}  (2006)  21--45.

\bibitem{Rokach2010}
L.~Rokach, Ensemble-based classifiers, {\em Artificial Intelligence Review}
  {\bf 33}  (2010)  1--39.

\bibitem{Leming2019}
M.~Leming and J.~Suckling, {Deep Learning on Brain Images in Autism: What Do
  Large Samples Reveal of Its Complexity?}, {\em Understanding the Brain
  Function and Emotions, \emph{Proceedings of the 8th International
  Work-Conference on the Interplay Between Natural and Artificial Computation,
  Part I}\/},  eds. J.~Ferrandez~Vicente, J.~Alvarez-Sanchez, F.~de~la
  Paz~Lopez, J.~Toledo~Moreo,  and H.~Adeli (Springer, 2019), pp. 389--402.

\end{thebibliography}
\end{multicols}
\end{document}